\documentstyle[preprint,aps]{revtex} 
\begin{document}
\title{Exponential Distributions in a Mechanical Model for Earthquakes}
\author{M. de Sousa Vieira\cite{email}}
\address{Departamento de Fisica, Universidade Federal do Rio Grande do Norte, 59072-970 Natal, RN, Brazil }
\address{Department of Electrical Engineering and Computer Sciences, University of California, Berkeley, CA 94720, USA.}
\maketitle
\begin{abstract}
We study statistical distributions in a  mechanical model for an 
earthquake fault introduced by Burridge and Knopoff [R. Burridge 
and L. Knopoff, {\sl Bull. Seismol. Soc. Am.} {\bf 57}, 341 (1967)]. 
Our investigations on the size (moment), time 
duration and number of blocks involved in an event show that exponential
distributions are found in a given range of the paramenter space. 
This occurs when the two kinds of springs present in the 
model have the same, or approximately the same, value for the elastic 
constants. 
Exponential 
distributions have also been seen recently in an experimental system to model 
earthquake-like dynamics [M. A. Rubio and J. Galeano, {\sl Phys. Rev. E} 
{\bf 50}, 1000 (1994)].      
\end{abstract} 
\narrowtext   
\section {Introduction} 
Systems that present stick-slip dynamics and scaling invariance have 
attracted considerable attention recently\cite {soc,train}. 
This was triggered, in part, by a seminal paper of Bak, Tang and 
Wiesenfeld, which showed that a class of systems presenting avalanches 
ou earthquake-like dynamics can naturally attain a critical state 
characterized by power-law distributions\cite{soc}. They 
denoted this phenomenon Self-Organized Criticality.  

Recently, two experimental studies in a continuous elastic system have 
been performed to model the stick-slip dynamics observed in 
earthquake-like phenomena. 
In one of them\cite{gollub}, a glass rod is pulled on top 
of a latex membrane. In this system the slipping events appear as 
detachment waves and a wide distribution of event sizes was observed. 
However, 
robust scaling behavior was not seen.

In the other system\cite{rubio}, the elastic medium consists of a 
transparent gel sheared between two coaxial circular cylinders. The 
inner cylinder is rotated at very low angular speed. Thus, this 
system closely resembles the physics of a spring-block model\cite{bk} with 
periodic boundary conditions. Due to the friction between the 
gel and the cylinders, stick-slip behavior is seen in the form of 
detachment waves.  
Rubio and Galeano were able to identify four regimes in this system,  
by varying the gel's rigidity, the rotor angular speed and the 
friction properties of the gel with the inner cylinder. 
The regimes identified were: (a) uniform slipping of the gel 
with respect to the rotor; (b) relaxation events, in which a big  
event starts to quickly propagates through the whole cell in an  
almost periodic way; (c) nearly periodic regimes of a soliton-like 
character and  (d) a regimes involving events with many 
size and time scales. In the latter case, 
they found exponential statistical 
distributions in amplitude, duration and separation time between events. 

With respect to theoretical  and numerical studies, one of the systems used  
to model the dynamics of earthquakes 
has been the model introduced by Burridge and Knopoff (BK) in 1967, and 
investigated extensively in recent publications \cite{cl,gio}. 
Although most of the studies on the Burridge-Knopoff model have been 
concentrated on the parameter region where a partial power-law (partial 
here means of limited size) of the distributions of event sizes 
is found, this regime in fact has not been 
seen yet in experimental studies of a homogeneous system.  The 
three first regimes (a), (b) and (c) found by Rubio and Galeano 
have been observed in numerical studies of the BK model. No reports so far exist of 
exponential distributions in the BK model.    

In this paper we show that exponential distributions are also   
seen in the BK model. This kind of distribution is observed 
when the two kinds of springs in the model have the 
same, or approximately the same, 
value for the  
elastic constants. 
They are present in quantities such as the time 
duration, number of blocks displaced, and total displacement (moment) of the 
blocks involved in an event. 

The paper is 
organized as follows: in the next section we describe the model; 
in the third section we study the statistical distributions and 
the different regimes that appear 
when the spring constants are varied; the last section is 
dedicated to the conclusions.  

\section{The Burridge-Knopoff Model}

The homogeneous version of the  Burridge and Knopoff model is shown in Fig. 1 
of Ref.~\cite{cl}.  
It consists of a one dimensional 
array of $N$ blocks, each of mass $m$, coupled by 
springs of constant $k_c$ to one another, and by  a 
spring of constant $k_p$ to a rigid pulling bar that moves at 
constant velocity $V$. In equilibrium, when all the springs are      
unstretched, adjacent blocks are separated by a distance $a$. The 
blocks rest upon a stationary surface, which provides a frictional 
force that impedes the motion of the blocks. In the  
version considered in Ref.\cite{cl} the friction is a decreasing 
function of the velocity, the same for all blocks. The equation 
of motion for the $j$th  mass when it is moving is  
\begin{equation}
m\ddot X_j=k_c(X_{j+1}-2X_j+X_{j-1})-k_p(X_j-Vt)-F(\dot X_j/V_f), \ \ \ \dot X_j\ne 0, 
%{{F_0}\over{1+|\dot X_j/V_f)|}}sgn(\dot X_j), \ \ \ \dot X_j\ne 0
%F(\dot X_j/V_f), \ \ \ \dot X_j\ne 0
\label{eq1}
\end{equation}
where $X_j$ denotes 
the displacement of the  block measured 
with respect to the position where the 
sum of the elastic 
forces on it is zero. 
The last term in Eq.~(\ref{eq1}) represents the  nonlinear and 
velocity dependent frictional force, which is given by 
%\begin{equation}
\[ F(\dot X_j/V_f)= \left\{ \begin{array} {cccc}   
                         F_0/(1+\dot X_j/V_f),& {\rm if}\ \dot X_j > 0\\
                        - \infty ,& {\rm if} \ \dot X_j < 0
                     \end{array}
\right. \]
\label{eq11}  
%\end{equation}
where 
$V_f$ is the characteristic velocity for the friction. 
Here, we do not allow backward motions and the friction force gets 
arbitrarily high to avoid it. This is just for computational 
convenience, and this does not change the main results presented here.  

As shown in  Ref. \cite{gio}, this model has five independent velocity 
scales. Two appear explicitly in Eq.~(\ref{eq1}): 
(i) the pulling velocity $V$ 
and (ii) the characteristic friction $V_f$. Two more can be defined 
in terms of the spring constants, the mass and $F_0$: 
(iii) $V_0=F_0/\sqrt{mk_p}$ and (iv) $V_l=(F_0/k_p)\sqrt{k_c/m}$.  
Here, $V_0$ and $V_l$ correspond to the maximum velocities of a single 
block held by a spring of constant $k_p$ and $k_c$, respectively, 
when it has displaced by the characteristic distance $F_0/k_p$ in the 
absence of friction. The fifth velocity is the sound velocity 
$V_s=a\sqrt{k_c/m}$, which  depends 
on the equilibrium spacing of the blocks $a$ and does not appear 
explicitly in Eq.~(\ref{eq1}). 
Depending on the relative sizes of these velocities, we 
would expect different behavior for the system.  

Following Ref.~\cite{cl}, we introduce different dimensionless 
variables    
\begin{equation}
U_j={{k_p}\over{F_0}}X_j, \ \ \ \ \tau=(m/k_p)^{1/2}t, 
\label{eq2}
\end{equation}
so that Eq.~(\ref{eq1}) can be written in dimensionless form 
\begin{equation}
\ddot U_j=\nu_l^2(U_{j+1}-2U_j+U_{j-1})-U_j+\nu \tau-\Phi(\dot U_j/\nu_f), 
%{{sgn(\dot U)}\over{1+|\dot U_j/\nu_f)|}}, \ \ \ \dot U_j\ne 0 
\label{eq3}
\end{equation}
where 
\[ \Phi (\dot U_j/\nu _f)= \left\{ \begin{array} {ccc} 
                         1/(1+\dot U_j/\nu _f),& {\rm if} \ \dot U_j > 0\\
                         - \infty ,& {\rm if} \ \dot U_j < 0
                     \end{array}
\right. \]
\label{eq12}
Now, $\nu _l=\sqrt{k_c/k_p}$, $\nu=V/V_0$, and $\nu_f=V_f/V_0$. Dots 
here represent derivatives with respect to the scaled time $\tau $. 
The sound velocity becomes $\nu _s=a' \sqrt{k_c/k_p}=a' \nu _l$, where 
$a'\equiv ak_p/k_c$. Without 
losing generality we can take the dimensionless length $a'\equiv 1$. 
Thus, the sound velocity $\nu _s$ becomes identical to $\nu _l$. 
%, and for periodic boundary conditions $U_0=U_N$.
The 
five velocities $V$, $V_f$, $V_l$, $V_s$ and $V_0$ have been transformed 
respectively into $\nu$, $\nu_f$, $\nu_l$, $\nu _s\equiv \nu _l$, and 
$\nu_0\equiv 1$. Consequently, the model has three relevant parameters, 
$\nu $, $\nu _f$ and $\nu _l\equiv \nu _s$, which  
completely determine the behavior of the 
system. 
In the case of open boundary condition, which is the one we consider here, 
we have 
$U_0=U_1$ and $U_{N+1}=U_N$. 
 
For fixed $\nu _l$ and $\nu $ (large $\nu _l$ and small $\nu $), a transition 
was reported in Ref.~\cite{gio} when  two velocities cross, namely, 
when $\nu _f=\nu_0\equiv 1$. For $\nu _f = \infty $ one finds that the motion 
of the system is continuous. No block ever stops. As  $\nu _f $ decreases, 
one sees small regions of stationary  blocks. When $\nu _f$ becomes 
less than $1$ these stationary (event-free)  regions begin to percolate 
across the entire system. The motion of the fault in this small $\nu _f$ 
region now occurs in abrupt large events. 
   
Here we report another transition, which occurs when the 
the  velocity $\nu _l$ is varied for fixed values of $\nu$ and $\nu _f$. 
We find that for $\nu _l$ less than a given value  the event sizes and 
event durations obey exponential distributions. 
Another  value is observed for $\nu _l$, above which the 
distributions present partial power-laws, namely, events that have 
size smaller than a critical value have power-law distributions, and events  
that are larger than this value obey different 
statistics. In the intermediate regime for $\nu _l$, the 
distributions are neither power-laws nor exponentials. 
As  $ \nu _l $ gets close to zero, the exponential distributions 
disappear, since in that limit the blocks are disconnected, and only 
one-block events are observed. 
Evidence of a different behavior when $\nu _l$ is varied was found in 
Ref.\cite{espanol}. However, no statistical distributions were 
calculated in that paper.   

In the numerical simulations of this letter we have generally  
started the system with the blocks at rest and with 
the sum of the elastic forces in each block equal to zero.  
Other initial conditions were considered and the 
results did not change. 
We have also considered periodic boundary conditions, and the results 
were the same (except for the few events involving the boundaries of the 
chain).  
Before we start to compile  statistics
we let the system evolve until it  
reaches a statistical 
stationary  state. 
Here we fix the pulling velocity and the 
characteristic velocity to the following values:  
$\nu =0.01$ and $\nu _f=1/6$. It is 
beyond the scope of this paper to do a detailed study on the 
effects of variations in the pulling velocity and in the characteristic speed.  
Studies on event distributions for varying $\nu _f$ and $\nu $ with 
fixed $\nu _l$ can 
be found in [6].  
We expect that the kind of transition we report here will be observed  
for other values of $\nu _f$ and $\nu $,   
and the value of $\nu _l$ where the transition occurs  
probably varies with the characteristic and pulling velocities.  

\section{Numerical Results}

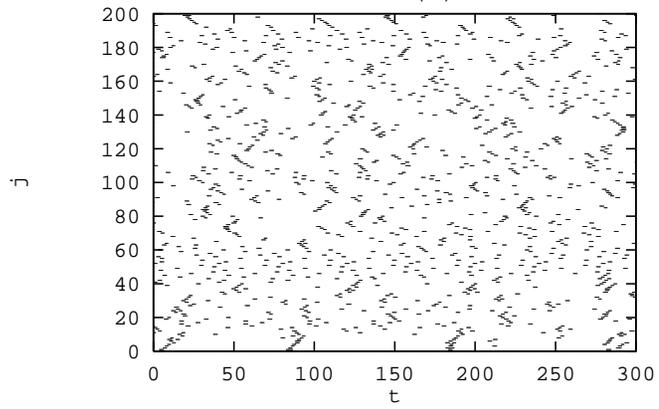
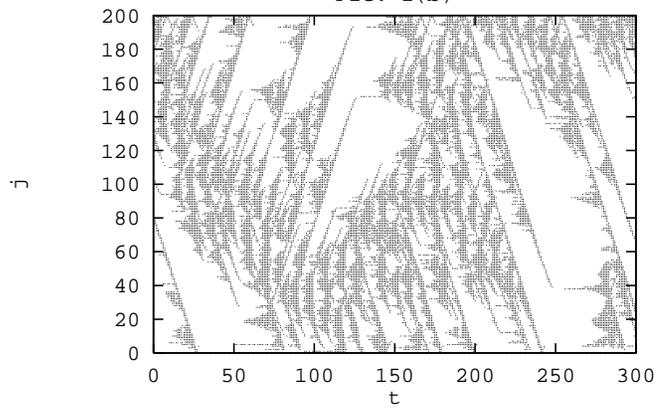
\begin{figure}
\caption[f2]{
Projections of the block velocities $\dot U_j$ onto the $j-\tau $ plane 
for a 200 block system with $\nu =0.01$, $\nu _f=1/6$, and (a) 
$\nu _l=1$, (b) $\nu _l=3$, and (c) $\nu _l=10$.}  
\label{f2}
\end{figure}

To illustrate the different behaviors of the system when $\nu _l$ is varied, we 
show in Fig. \ref{f2} projections of the block velocities $\dot U_j$ onto the 
$j-\tau$ plane. A black dot in the figure means that the $j$-block is moving at 
time $\tau $ and 
a white one, that it is at rest. We show three different cases with 
$\nu _l=1, 3, 10$. It is clear that a 
transition occurs as $\nu _l$ is changed. For small $\nu _l$, we 
see clusters of moving blocks, which involve only a small number of masses. 
As $\nu _l$ is increased, one finds clusters involving a small number 
of blocks, as well as clusters involving a large number of masses. These 
large ruptures appear $\vee$ shaped and travel with the  
sound speed $\nu _l$.  

Now we investigate the distributions of  moment and time duration 
for different values of $\nu _l$.
The moment of an earthquake is 
a measure of its size, and in dimensionless 
units it is defined as  
$m=\sum_j \delta U_j$, 
where $\delta U_j$ is the displacement of the $j$-th block and  
the sum is over the blocks displaced during the event. 
The distribution $r (m)$ is the number of events with magnitude $m$ 
divided by the number of blocks $N$  in the chain and by the
total time of dynamical evolution. 
For  $\nu _l \lesssim 2$  we find that $r (m)$ is governed by an  
exponential function,  namely, 
$r (m) \sim \exp(m/m_c)$, where $m_c$ is a characteristic moment. 
The result of $r (m)$ with $\nu _l=1$  
is shown in Fig.~\ref{f4}(a) as a solid line.    
We have also found that 
the characteristic moment decreases as $\nu _l$ becomes
smaller, since when $\nu _l=0$ the blocks become disconnected and only 
events involving one block are observed.
For intermediate values of $\nu _l$ ($2 \lesssim \nu _l \lesssim 4$)  ones sees a regime 
for $r (m)$ which is neither power-law nor exponential behavior. 
This is shown in Fig. \ref{f4}(b) where the solid line represents  $r (m)$ for
$\nu _l=3$.
For large $\nu _l$ ($\nu _l \gtrsim 4$)
one sees the appearance of a partial power-law distribution. 
This 
was the parameter region investigated by  
Carlson and Langer\cite{cl}. 
Beyond the 
power-law cutoff one sees a bump in the distributions, which 
increases in height as $\nu _l$ becomes larger. We have found that the events that do not 
belong to the scaling region become more and more frequent as $\nu _l$ 
is increased. Thus, in the limit of very large $\nu _l$ the power-law 
tends to disappear.  
In Fig. \ref{f4}(c) 
the solid line represents $r (m)$ for $\nu _l=10$. 

For the distributions of time durations $\delta \tau$ of the events 
we find qualitatively 
the same kind of behavior as found for the distributions of moments. 
For $\nu _l \lesssim 2$ we get an exponential function,     
$r (\delta \tau) \sim \exp(\delta \tau /\delta \tau _c)$. 
This is shown in 
Fig.~\ref{f4}(a) for $\nu _l=1$ as a dashed line. 
Again, we have found that the characteristic time duration 
$\delta \tau _c$ decreases as $\nu _l$ gets
smaller.
For $\nu _l \gtrsim 4 $ one sees partial power-law distributions, as shown 
in Fig.~\ref{f4}(c)  for $\nu _l=10$ as a dashed line.  
For intermediate values of $\nu _l$ ($2 \lesssim \nu _l \lesssim 4$) the 
distributions are neither exponential nor power-law ones, as we 
show in Fig.~\ref{f4}(b) for $\nu _l=3$ (dashed line). 

We have also studied the statistical distributions $r (n)$ of an event 
involving $n$
blocks. Again, the same kind of behavior described above is seen, namely, 
exponential distributions for $\nu _l \lesssim 2$, partial power-laws for 
$\nu _l \gtrsim 4$, and a regime without any scaling features for intermediate 
values of $\nu _l$.  

\section {Conclusions} 
We have found exponential distributions for the sizes and time 
durations of the events in the 
Burridge-Knopoff model for earthquakes when the two kinds of springs have 
the same (or approximately the same) elastic constants. In this situation 
the velocity $\nu _l$ and the sound velocity 
are equal (or approximately equal) to one. 
For intermediate values of $\nu _l$, the 
distributions are neither exponential nor power-law.     
For large $\nu _l$, partial power-laws (namely, a scaling region of 
limited size) are observed. 
Since in the experimental system studied by Rubio and Galeano\cite{rubio} 
the elastic gel is homogeneous, 
this implies that the exponential distributions 
they found are consistent 
with our results for the case in which the  
springs of the BK model have identical elastic constants.  
 
%\acknowledgments
%I would like to thank the Brazilian agency CNPq for the financial 
%support, and  Profs. L. Lucena and L. R. Silva for the warm hospitality at  
%the Universidade Federal do Rio Grande do Norte.  

\begin{figure}
\caption[f4]{
Statistical distribution $r (m) $ of the moment of the events 
(solid lines) and the distribution $r (\delta \tau) $ of the time 
durations  of the events (dashed lines)  
for (a) $\nu _l=1$,  (b) $\nu _l=3$ and (c) 
$\nu _l=10$. 
In all cases, $N=300$,  $\nu =0.01$ and $\nu _f=1/6$.}  
\label{f4}
\end{figure}


\begin{references}
\bibitem[*] {email} Present Address: Dept. of EECS - Cory Hall, University  
of California, Berkeley, CA 94720. Electronic address: mariav@eecs.berkeley.edu

\bibitem {soc} P. Bak, C. Tang, and K. Wiesenfeld, {\sl Phys. Rev. Lett.}
{\bf 59}, 381 (1987); {\sl Phys. Rev. A} {\bf 38}, 364 (1988).

\bibitem{train} M. de Sousa Vieira, {\sl Phys. Rev. A} {\bf 46}, 6288 (1992). 

\bibitem {gollub} D. P. Vallete and J. P. Gollub, {\sl Phys. Rev. E} {\bf 47}, 
820 (1993). 

\bibitem {rubio} M. A. Rubio and J. Galeano, {\sl Phys. Rev. E} 
{\bf 50}, 1000 (1994). 

\bibitem {bk} R. Burridge and L. Knopoff, {\sl Bull. Seismol. Soc. Am.} 
{\bf 57}, 
341 (1967).

\bibitem {cl} J. M. Carlson and J. S. Langer, {\sl Phys. Rev. Lett.} {\bf 62}, 2632 (1989); {\sl Phys. Rev. A} {\bf 40}, 6470 (1989).

\bibitem {gio} G. Vasconcelos, M. de Sousa Vieira and S. R. Nagel, {\sl Physica 
A} {\bf 191}, 69 (1992); M. de Sousa Vieira, G. Vasconcelos and S. R. Nagel, 
{\sl Phys. Rev. E} {\bf 47},  R2221 (1993).  

\bibitem {espanol} P. Espa\~ nol, {\sl Phys. Rev. E} {\bf 50}, 227 (1994). 

%\bibitem {roux} J. Schmittbuhl, J-P. Vilotte and S. Roux,
%{\sl Europhys. Lett.} {\bf 21} (1993) 375.

\bibitem {fric} M. de Sousa Vieira, H. J. Herrmann, {\sl Phys. Rev. E} 
{\bf 49} 4534 (1994).  


\end{references}
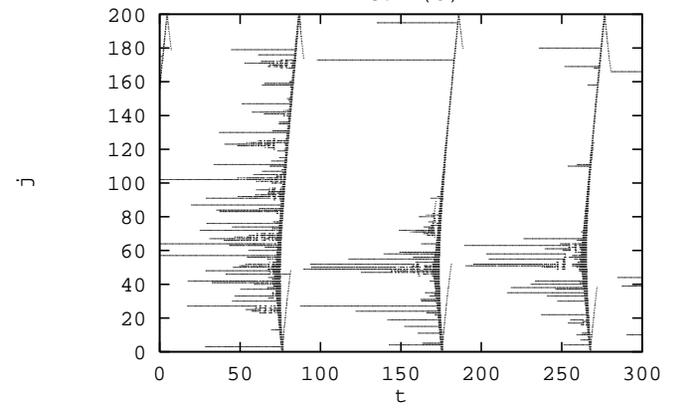
\end{document}